\documentclass[journal]{IEEEtran}
\ifCLASSINFOpdf
\else
\fi
\usepackage{epsfig}

\usepackage[cmex10]{amsmath}
\begin{document}
%
\title{Fountain Codes and Invertible Matrices}
%
%
%

\author{Mikko~Malinen,~\IEEEmembership{Student~Member,~IEEE.}}
\maketitle

\begin{abstract}
This paper deals with Fountain codes, and especially with their encoding matrices, which are required here to be invertible. A result is stated that an encoding matrix induces a permutation. Also, a result is that encoding matrices form a group with multiplication operation. An encoding is a transformation, which reduces the entropy of an initially high-entropy input vector. A special encoding matrix, with which the entropy reduction is more effective than with matrices created by the Ideal Soliton distribution is formed. Experimental results with entropy reduction are shown. 
\end{abstract}

\begin{IEEEkeywords}
Entropy-reducing transformation, Fountain codes, group, Ideal Soliton distribution, permutation.
\end{IEEEkeywords}

%
\IEEEpeerreviewmaketitle

\newtheorem{Conj}{Conjecture}

\section{Introduction}

%
%
%
%
\IEEEPARstart{F}{ountain} codes were first mentioned in $[1]$. LT-codes $[3]$ are first practical Fountain codes. In Fountain codes we have $k$ input symbols and $n$ output symbols. We call an encoding graph a bipartite graph where on the other side are the input symbols and on the other side the output symbols. There are edges which mark the connections between the input symbols and the output symbols. We call the degree of an output symbol the number of input symbols the output symbol is connected to. The degree distribution $\rho (i)$ tells the probability that an output symbol has $i$ connections. We call an encoding matrix $R$ a $k\times k$ matrix where an element $r_{lm}$ is 1 if the $m$:th input symbol affects (has a connection) to the $l$:th output symbol and 0 if it has not. Generally, $R$ may have rank $< k$, but here we restrict the treatment to matrices which have full rank, i.e. all the rows are linearly independent. These matrices are invertible. This way the number of output symbols $n$ equals the number of input symbols $k$. Fountain codes for which the matrix $R$ is of full rank are always decodable.\\
\ \\
The output bits are calculated by
$$y = Rx,$$
where $y$ is output bits in vector form, $x$ is input bits in vector form and the multiplication is done modulo 2 as is the idea in Fountain coding. The decoding is done by
$$x = R^{-1}y.$$

\section{Encoding matrix induces a permutation}

Our first result (Result 1) is that when there are two different inputs $x^{(1)}$ and $x^{(2)}$, the two outputs $y^{(1)}$ and $y^{(2)}$ are always different. This is due to the fact that when decoding $y^{(i)}$ we end up to an unique $x^{(i)}$.\\
\ \\
From the Result 1 follows the next result: By multiplying an input vector $x$ several times by the encoding matrix, we end up back to $x$ at some point. Each multiplication results to different output until we end up to $x$. Depending on the choice of the initial input we end up to different cycles. Thus, in principle, we could make decoding of an output by multiplying the output, length of the cycle - 1 times. Of course, we have to know the length of the cycle. If we number the different length $k$ vectors by $1,2,...,2^k$ we can say that an encoding matrix $R$ induces an unique permutation. We can use the list presentation of a permutation $[2]$, pp. 52-64, to express the permutation. From the list presentation can be seen that there are $s!$ different possible permutations of $s$ elements. This could be used as the upper bound for the number of different encoding matrices. However, it turns out that $2^k!$ (we have $2^k$ elements) grows faster than the total number of different 0-1 matrices $2^{(k^2)}$ (in a $k\times k$ matrix there are $k^2$ elements). Thus this upper bound is practically useless. We may also conclude that invertible 0-1 matrices of size $k\times k$ do not induce all possible permutations of size $2^k$.

\section{Encoding matrices form a group}
One result is that encoding matrices (invertible 0-1 matrices) form a group with modulo 2 multiplication operation. Namely, two such matrices multiplied is also such a matrix. There is a unit element, a unit matrix. Also, the inverse element is the inverse matrix. And the multiplication is associative.

\section{Entropy-reducing transformations}

\begin{figure} 
\begin{center}
\epsfig{file=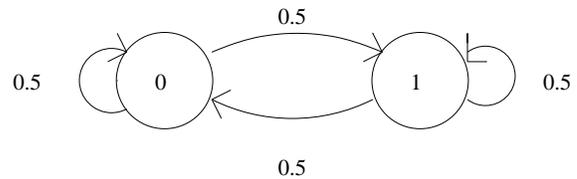,width=8cm}
\ \\
\caption{A stochastic process defined by a Markov chain.} \label{fig:markov}
\end{center}
\end{figure}

Next, we come to entropy considerations. We state a result, that when all different input vectors are multiplied by an encoding matrix, the entropy remains the same on average. This is due to the fact that if the entropy would change, we could compress the input vector below what the initial entropy indicates. Even if the entropy increased on the average, we could reduce the entropy by using the inverted matrix. However, according to our experimental results, if the entropy is "big", i.e., the number of 0's and 1's are near the same, the entropy decreases on the average when multiplied by an encoding matrix. This we have calculated on a 8 bit long input and the Ideal Soliton degree distribution as described in $[3]$. As we shall see later, we can construct a special encoding matrix with almost all distribution weigth on degree 2 with which the reduction in entropy is demonstrated also on large input lengths. For the first mentioned 8 bit long input, with 4 zeros and 4 ones we give the probabilities of zero in output in Table~\ref{ta:tableone} ($\rho (i)$ is the Ideal Soliton degree distribution):
\begin{table}[h]
\begin{center}
\begin{tabular}{|l|l|l|} 
\hline
degree $i$ & $\rho (i)$ & Probability of zero in output\\
\hline
1 & 1/8 & 0.5\\
2 & 1/2 & 0.42857\\
3 & 1/6 & 0.5\\
4 & 1/12 & 0.52857\\
5 & 1/20 & 0.5\\
6 & 1/30 & 0.42857\\
7 & 1/42 & 0.5\\
8 & 1/56 & 1\\
\hline
\end{tabular}
\ \\
\ \\
\caption{Ideal Soliton degree probabilities and probabilities of zero in output for a 8-bit length input with equal number of zeros and ones} \label{ta:tableone}
\end{center}
\end{table}
\ \\
The two last columns multiplied with each other and summed results in 0.47321 probability of zero in output. This is lower than the 0.5 probability at the input. One may think that there is some other degree distribution with even better reduction in entropy or that there exists some degree distribution that gives the best reduction. It can be seen from the Table~\ref{ta:tableone} that degree $i=2$ gives the lowest probability of zero in the output, 0.42857. We used this degree and formed a special invertible encoding matrix $R$:
$$r_{lm} = 1, \hspace{2mm} \text{when} \hspace{2mm} m=l \hspace{2mm} \text{or} \hspace{2mm} m = l+1$$
$$r_{lm} = 0, \hspace{2mm} \text{otherwise}.$$
For example, a $4\times 4$ encoding matrix $R$ would be: 
$$R =
\begin{bmatrix}
1&1&0&0\\
0&1&1&0\\
0&0&1&1\\
0&0&0&1
\end{bmatrix}
.
$$
There is a single 1 in the last row at the last column. These matrices are invertible and thus suitable for encoding and decoding. We ran simulations with input length 30204 and 61408. We used different initial proportions of zeros and ones in input. We calculated the effect of the entropy change to the optimal representation of the bits.
\begin{figure} 
\begin{center}
\epsfig{file=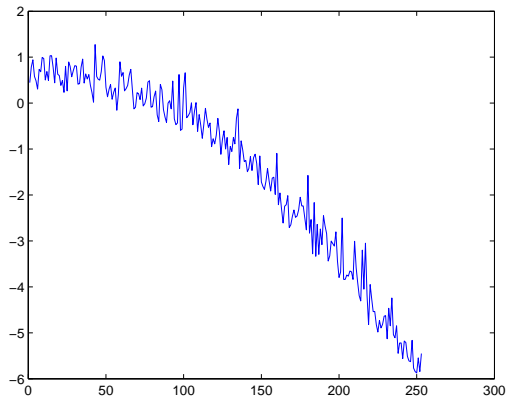,width=8cm}
\ \\
\caption{Saving in bits (vertical direction) when transformation is applied to input of length 30204 with different reduced number of ones in the input (horizontal direction)} \label{fig:saving}
\end{center}
\end{figure}
In Figure~\ref{fig:saving} we see the saving in bits as the function of uneveness of 0's and 1's at the input. The input vector length is 30204 bits. At the left border there are 0 ones less than zeros in input. Near the right border there are 250 ones less in input than in the left border. With each horizontal value ten simulations were made, with randomly placed ones, and the average was taken. We see that for horizontal values 0..90 the needed number of bits in representing the output is lower than the needed number of bits needed in representing the input. The reduction is at most near 1 bit. With larger horizontal values the entropy increases and more bits are needed to represent the output. In principle, we got similar results with input vector length 61408. The space saving is still at most less than one bit on average. The results rise some conjectures which we present here.
\begin{Conj}
The space saving using this transformation is on average less than one bit, even when applied to high-entropy input.
\end{Conj}
Let us consider a stochastic process defined by the Markov chain in Figure~\ref{fig:markov}.
The average entropy of a finite length realization of this process can be calculated with the help of a binomial probability density function multiplied by entropy function. It would be tempting to think that the average entropy equals to the entropy calculated from the proportions of 0's and 1's when the number of ones is decreased by the standard deviation of the associated binomial distribution. The standard deviation of a binomial distribution is $\sigma = \sqrt{np(1-p)}$. But, according to our calculations, this is not the case. These entropies differ. 
\begin{Conj}
The transformation described above can not reduce on average the entropy of a realization of the stochastic process described above below the average entropy of the same process.
\end{Conj}

\section{Conclusions}
Fountain codes and especially their invertible encoding matrices were dealt with. Encoding matrices induces a unique permutation on $k$ bit strings. Encoding matrices with multiplication operation form a group. Encoding matrices sampled from the Ideal Soliton degree distribution reduce entropy at least in a special case. A special transformation matrix model was formed and different input vectors were used in simulations which showed reduction in entropy when the initial entropy was high. Although some entropy reduction was possible, it was on average less than one bit from tens of thousands of bits, so in practice this this method may not be applicable. However, more research is still needed to show if Conjecture 1 holds.

\section*{References}

\ \\
$[1]$ J. Byers, M. Luby, M. Mitzenmacher, and A. Rege, "A digital fountain approach to reliable distribution of bulk data", in \emph{Proc. ACM SIGCOMM '98}, Vancouver, BC, Canada, Jan. 1998, pp. 56-67\\
\ \\
$[2]$ D. L. Kreher and D. R. Stinson, \emph{Combinatorial Algorithms, Generation, Enumeration and Search}, CRC Press LLC, Boca Raton, FL, 1999\\
\ \\
$[3]$ M. Luby, "LT-codes", in \emph{Proc. 43rd Annu. IEEE Symp. Foundations of Computer Science (FOCS)}, Vancouver, BC, Canada, Nov. 2002, pp. 271-280\\

\end{document}